\documentclass[prd,reprint]{revtex4-1}

\usepackage{amssymb} 
\usepackage{amsmath}
\usepackage{xfrac}
\usepackage{bm}

\newcommand{\hf}{\frac{1}{2}}
\newcommand{\fb}[2]{\left(\frac{#1}{#2}\right)}

\begin{document}

\hphantom\\
\begin{flushright}
${}$\\[-2cm]
MAN/HEP/2018/02\\
June 2018
\end{flushright}

\title{\Large{The Eisenhart Lift for Field Theories}}

\author{Kieran Finn, Sotirios Karamitsos and Apostolos Pilaftsis} 

\affiliation{School of Physics and Astronomy, University of Manchester, Manchester
 M13 9PL, United Kingdom}


\begin{abstract}

We present the Eisenhart-lift formalism in which the dynamics of a system that evolves under the influence of a conservative force is equivalent to that of a free system embedded in a curved manifold with one additional generalised coordinate. As an illustrative example in Classical Mechanics, we~apply this formalism to simple harmonic motion. We extend the Eisenhart lift to homogeneous field theories by adding one new field. Unlike an auxiliary field, this field is fully dynamical and is therefore termed \emph{fictitious}. We~show that the Noether symmetries of a theory with a potential are solutions of the Killing equations in the lifted field space. We generalise this approach to field theories in four and higher spacetime dimensions by virtue of a mixed vielbein that links the field space and spacetime. Possible applications of the extended Eisenhart-lift formalism including the gauge hierarchy problem and the initial conditions problem in inflation are briefly discussed.

\end{abstract}

\maketitle

\section{Introduction}
\label{sec:intro}
In Newtonian mechanics, fictitious forces arise in non-inertial frames of reference. In an inertial frame, a particle moves according to Newton's laws of motion \cite{newton1687}. However, if we transform to a non-inertial frame, it is necessary to include additional forces to correctly describe the trajectory of the particle. Examples include the centrifugal and Coriolis forces that appear in a rotating reference frame such as the Earth.

These fictitious forces are always proportional to the inertial mass of the particle on which they act. Since gravity is proportional to gravitational mass, one may wonder whether the principle of equivalence implies that gravity is a fictitious force as well. Based on this idea, Einstein showed in his theory of General Relativity${}$~(GR)~\cite{Einstein:1915ca} that the effects of gravity actually arise from the curvature of spacetime. If we treat spacetime as flat, as we do in Newtonian mechanics, we must add gravity as a fictitious force to account for this curvature${}$.

Soon after the development of GR, the question of whether it is possible to incorporate other forces in a similar\- fashion was raised. One such attempt was Kaluza--Klein theory \cite{Kaluza:1921tu,Klein:1926tv,Overduin:1998pn}, which considers a five-dimensional space with a compactified fifth coordinate. It was shown that observers living in a four-dimensional subspace of this theory would experience some of the additional degrees of freedom in the metric as a vector-boson field satisfying Maxwell's equations. Hence, electro\-magnetism can be regarded as a fictitious force, since it arises only from the requirement that particles follow geodesics in the five-dimensional space.

A question that naturally follows is whether this geometric approach can be extended to all forces. This question was asked by Eisenhart~\cite{eisenhart},  who demonstrated that the trajectory of a particle subject to a potential can be described as a geodesic on a higher dimensional manifold. Therefore, the force that arises from the gradient of the potential is fictitious, just as electro\-magnetism is in Kaluza--Klein theory. This formalism is known as the \emph{Eisenhart lift}, and has since been extensively studied in the context of integrable systems in \mbox{Classical Mechanics}~\cite{Duval:1984cj,Minguzzi:2006gq,Cariglia:2016oft,Cariglia:2018mos}; a pedagogical introduction can be found in \cite{Cariglia:2015bla}.

In this paper we extend the applicability of the Eisenhart lift to scalar field theories. We are motivated by the fact that a theory with multiple homogeneous scalar fields (such as leading-order multifield inflation \cite{Sasaki:1998ug,Gordon:2000hv,Seery:2005gb}) has the same mathematical structure as that of a classical particle moving under the influence of a potential. We can thus define a \emph{field space} \cite{Arroja:2008yy,Langlois:2008qf,GrootNibbelink:2000vx,GrootNibbelink:2001qt,
vanTent:2003mn,Kaiser:2013sna,Reuter:2015rta,Karamitsos:2017elm}, as introduced in Section~\ref{sec:field space}, such that the evolution of the system will be a trajectory in that space.

If the theory has no potential, then this trajectory will be a geodesic of the field space. Adding a potential term to the theory causes the trajectory to deviate from that geodesic. However, using the Eisenhart lift, we can reproduce the effects of this potential by defining a higher dimensional field space with one additional field, such that the system does follow a geodesic in this extended space. Unlike an auxiliary field, this field is fully dynamical and as such, we call it \emph{fictitious}.

The one-to-one correspondence between a scalar field theory and a particle moving in curved space is lost when the fields depend on more than one coordinate. Thus we must extend the Eisenhart-lift formalism in order to deal with non-homogeneous field theories. To this end, we add a fictitious vector field to the theory that links the field space and spacetime. In this way, we are able to construct a purely kinetic Lagrangian with an extended field space that yields the original equations of motion (EoMs) for an arbitrary scalar field theory when the fictitious degrees of freedom are projected out.

The paper is laid out as follows. We first outline the Eisenhart lift in Section${}$~\ref{sec:classical} by showing how a conservative force acting on a classical\- particle may be interpreted as a fictitious force. In Section${}$~\ref{sec:SHM}, we illuminate this formalism by studying the simple harmonic oscillator as a concrete example. Next, in Section~\ref{sec:field space}, we introduce the notion of a field space in theories with multiple scalar fields and show how, for homogeneous fields, we can use the same formalism to recover the effects of any potential term with the aid of a fictitious${}$ field. In Section~\ref{sec:Killing}, we show how the Noether symmetries of the theory become solutions of Killing's equation in this new extended field space. In Section~\ref{sec:4d}, we generalise the Eisenhart lift to field theories in higher-dimensional curved spacetime. Finally${}$, we conclude in Section~\ref{sec:discuss} by discussing our findings and presenting possible applications of employing the Eisenhart-lifted field space.

Throughout this paper we use natural units where ${c=\hbar=1}$.

\section{The Eisenhart Lift}
\label{sec:classical}
Let us consider a particle with mass $m$ moving in $n$ dimensions under the influence of a potential $V({\bf x})$. Here we denote the coordinates of the space individually by~$x_i$ (with $1\leq i \leq n$) and collectively by ${\bf x}$. Such a particle has the following Lagrangian:
\begin{equation} 
L\ =\ \frac{1}{2}  m\sum_{i=1}^n \dot x_i^2\: -\: V({\bf x})\;,
\end{equation}
where the overdot denotes differentiation with respect to time $t$. The evolution of the system is governed by Newton's second law \cite{newton1687}:
\begin{align} \label{eq:eomnewton}
m \ddot x_i\ =\  -\,V_{,i}\;,
\end{align}
where ${,i}$ denotes differentiation with respect to $x_i$. We see that the derivative of the potential yields a conservative force that causes the trajectory of the particle to deviate from a geodesic (a straight line in this case). However, as Eisenhart showed, this force is fictitious. This means that the true trajectory of the particle is a geodesic in an \mbox{$(n\!+\!1)$--dimensional} coordinate space and the potential term is only required because we have restricted our attention to an $n$ dimensional subspace.

To make the Eisenhart lift explicit, let us follow his construction and extend the space by adding a new coordinate $y$ and consider a new Lagrangian
\begin{equation}\label{eq:lagchi}
L\ =\ \frac{1}{2}  m\sum_{i=1}^n \dot x_i^2\: +\: \frac{1}{2}\frac{M^2}{V({\bf x})}\dot y^2\, ,
\end{equation}
where $M$ is an arbitrary mass scale introduced to keep the dimensions consistent. This is the Lagrangian of a \emph{free} particle on a curved manifold with coordinates ${x_I\in\{x_i,y\}}$ and a metric
\begin{equation}
g_{IJ}\ =\ \begin{pmatrix}
\delta_{ij}&0\\
0&\dfrac{M^2}{mV}
\end{pmatrix}.
\end{equation}
This implies that the particle will follow a geodesic of that manifold.

We will now show that geodesics of the above manifold endowed with the metric $g_{IJ}$ reduce to the EoMs~\eqref{eq:eomnewton} when projected down to an $n$-dimensional submanifold. The EoMs  obtained from the Lagrangian~\eqref{eq:lagchi} are found to be
\begin{equation}
m \ddot x_i\ =\ -\, \frac{M^2}{2}\frac{V_{,i}}{V^2}\: \dot y^2\; ,\hspace{2em}
\frac{d}{dt} \left( \frac{\dot y}{V({\bf x})}\right)\ =\ 0\;.\label{eq:classical eoms}
\end{equation}
The solution to the latter of these equations is ${{\dot y} = A V/M}$, where $A$ is a constant determined by the initial\- conditions. If we restrict ourselves to solutions for which $A^2 = 2$, we find that after substitution into the first equation of~\eqref{eq:classical eoms}, we recover the original EoMs~\eqref{eq:eomnewton}. 
 
The requirement of $A^2=2$ has a simple geometric interpretation. The motion of a particle in $n\!+\!1$ dimensions follows a geodesic. However, we have some freedom in how to parametrise this trajectory. We have to choose which parameter within the affine class should be identified with the time~$t$. The choice we make determines the value of $A$. Only $A^2=2$ will give us the EoMs \eqref{eq:eomnewton}. If we choose $A^2\neq2$, the system will either evolve in slow motion${}$~($A^2<2$) or fast-forward ($A^2>2$). We will see a specific example of this property in the next section.

Thus, a manifold has been found whose geodesics correspond to the evolution of a system subject to a conservative force. Since no particular form of the potential term has been assumed, this approach will work for any conservative force. Hence, it has been shown that all conservative forces are fictitious and can be described by the geometry of a higher dimensional manifold. 

\section{Example: Simple Harmonic Motion}
\label{sec:SHM}
In order to give a concrete example of the Eisenhart-lift formalism, we consider the simple harmonic oscillator. This system is governed by the Lagrangian
\begin{equation}\label{eq:SHM lagrangian}
L\ =\ \hf m \dot{x}^2\: -\: \hf k x^2\;.
\end{equation}
Using the formalism described in Section~\ref{sec:classical}, we define a new Lagrangian with an additional coordinate $y$,
\begin{equation}
L\ =\ \hf m\dot{x}^2\: +\: \frac{1}{4}\frac{k}{m^4 x^2}\dot{y}^2\;,\label{eq:SHM new lagrangian}
\end{equation}
where we have chosen the arbitrary mass scale to be: ${M= k/(2\,m^2)}$, so that the new Lagrangian reduces to~\eqref{eq:SHM lagrangian} in the limit $k\to 0$.

We now proceed by finding and solving the EoMs for this system. The Euler--Lagrange equations 
for this system\- are given by
\begin{equation}\label{eq:SHM EOM}
m\ddot{x}\ =\ -\,\frac{k}{m^4x^3}\:\dot{y}^2\;,\hspace{2em}
\frac{d}{dt}\fb{\dot{y}}{x^2}\ =\ 0\;.
\end{equation}
The solution to the second equation is
\begin{equation}\label{eq:SHM chidot sol}
\dot{y}\ =\ A m^2 x^2,
\end{equation}
where $A$ is a constant. Substituting this solution into the first equation in \eqref{eq:SHM EOM} gives
\begin{equation}
\ddot{x}\ =\ -\frac{A^2}{2}\frac{k}{m}x\;.\label{eq:SHM xddot sol}
\end{equation}
This is the equation of simple harmonic motion and has the following\- solution:
\begin{equation}\label{eq:xsol}
x(t)\ =\ x_0\,\cos\left[\frac{A}{\sqrt{2}}\omega (t-t_0)\right],
\end{equation}
where $\omega=\sqrt{k/m}$, and $x_0$ and $t_0$ are parameters set by the initial conditions. We can now substitute~\eqref{eq:xsol} into~\eqref{eq:SHM chidot sol} and solve the latter by direct integration to find
\begin{equation}
\label{eq:ysol}
y(t)\ =\ y_0\: +\: \frac{A}{\sqrt{2}}\frac{m^2x_0^2}{2} t\: +\: \frac{m^2x_0^2}{4\omega}\sin\left[\sqrt{2}A\omega(t-t_0)\right]\,,
\end{equation}
where $y_0$ is a constant of integration.

The solution given by \eqref{eq:xsol} and \eqref{eq:ysol} has four free parameters. The amplitude is set by $x_0$ and the phase is set by $t_0$ as usual. The parameter $y_0$ controls only the initial value of $y$ and hence is irrelevant as our theory is shift-symmetric in $y$. This leaves $A$, which enters the EoMs only as a multiplicative factor of $t$. Thus, any dependence on $A$ can be removed by simply rescaling $t$ as explained in the previous section. In other words, $A$ dictates how quickly the system evolves. If we choose $A=\sqrt{2}$, $x$ obeys the same EoM as it would for the Lagrangian~\eqref{eq:SHM lagrangian}.

\begin{figure}
\includegraphics[width=\columnwidth]{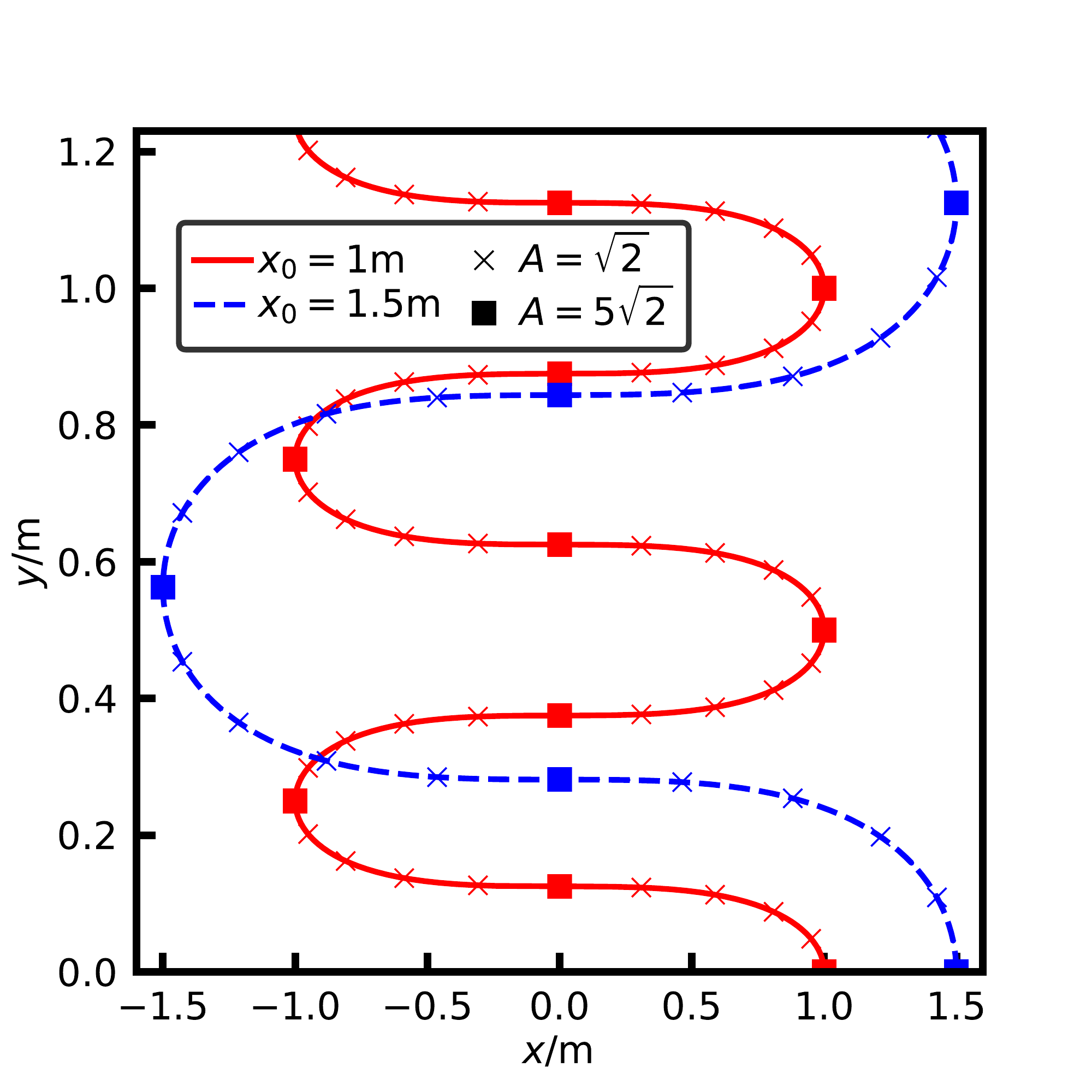}
\caption{Motion of the simple harmonic oscillator in the extended space with $\omega=2\pi \,\rm{s}^{-1}$ and $m=1\,\rm{kg}$ and $y_0=t_0=0$. The solid (red) line and dashed (blue) line indicate trajectories with oscillation amplitudes $x_0=1\,\rm{m}$ and $x_0=1.5\,\rm{m}$, respectively. The markers are placed at equal time intervals of $\Delta t=0.05\,\rm{s}$ with crosses for $A=\sqrt{2}$ and squares for $A=5\sqrt{2}$. They show how varying $A$ does not change the trajectory, only how quickly it is traversed.
\label{fig:SHM}}
\end{figure}

Figure~\ref{fig:SHM} displays the trajectory of the simple harmonic\- oscillator in the extended manifold with parameters $y_0=t_0=0$, $m=1\,$kg and $\omega=2\pi\, \rm{s}^{-1}$ in SI units. Trajectories are shown for $x_0=1\,\rm{m}$ (solid red) and $x_0=1.5\,\rm{m}$ (dashed blue) that illustrate the effect of changing the amplitude of oscillation. We also demonstrate the effect of $A$ by placing crosses and squares for $A=\sqrt{2}$ and $A=5\sqrt{2}$, respectively, at equal time intervals of $0.05\,\rm{s}$. As expected, the trajectory is unaffected by varying $A$, but the speed at which the trajectory is traversed increases when we change $A$ from $\sqrt{2}$ to $5\sqrt{2}$. Notice that there is $0.5\,\rm{s}$ between peak and trough when $A=\sqrt{2}$ ($0.1\,\rm{s}$ when $A=5\sqrt{2}$) independently of the value of $x_0$. Thus, we have demonstrated that the period of oscillation remains independent of the amplitude.

\section{Field space}
\label{sec:field space}
We will now move from Classical Mechanics to Field Theory. We will initially focus on homogeneous field theories by outlining the description of the classical field space. We consider a theory of $N$ scalar fields $\varphi^i(t)$ (collectively denoted by $\bm{\varphi}$) with an arbitrary quadratic kinetic term but no potential term. Such a theory can be described by the following Lagrangian:
\begin{equation}\label{eq:free lagrangian}
{\cal L}\ =\ \hf k_{ij}(\bm{\varphi})\;\dot{\varphi}^i\dot{\varphi}^j\;,
\end{equation}
where the indices $i$ and $j$ run from $1$ to $N$. The EoMs for the fields~$\varphi^i$ are given by
\begin{equation}\label{eq:geodesic}
\ddot{\varphi}^i \:  +\:   \hf k^{il}\Big(k_{jl,k}  +   k_{kl,j}   -  k_{jk,l}\Big)\,
\dot{\varphi}^j\dot{\varphi}^k\  =\ 0\;,
\end{equation}
where $,i$ now indicates a partial derivative with respect to the field $\varphi^i$ and $k^{ij}$ is the inverse of $k_{ij}$ satisfying $k^{il}k_{lj}=\delta^i_j$. We notice that if we interpret $k_{ij}$ as the \emph{field space metric}, \eqref{eq:geodesic} becomes simply the geodesic equation
\begin{equation}
\frac{\partial^2\varphi^i}{\partial u^2}\:  +\:  \Gamma^i_{jk}\,\frac{\partial \varphi^j}{\partial u}\frac{\partial \varphi^k}{\partial u} \  =\  0\;,\label{eq:geodesic u}
\end{equation}
where we identify the affine parameter~$u$ with time~$t$ and the Christoffel symbols
\begin{equation}
\Gamma^i_{jk}\ =\ \hf k^{il}\,\Big(k_{jl,k}+k_{kl,j}-k_{jk,l}\Big)
\end{equation}
play the role of the affine connection.

This latter formulation leads to the following inter\-pretation of the theory. We define an $N$-dimensional field space with coordinates~$\varphi^i$ and equip it with a metric $k_{ij}$. Thus, our field configuration at a given time corresponds to a ``particle'' that traverses the field space, and the evolution of our theory corresponds to its trajectory~\mbox{\cite{Sasaki:1998ug,Gordon:2000hv,Seery:2005gb,Arroja:2008yy,
Langlois:2008qf,GrootNibbelink:2000vx,GrootNibbelink:2001qt,vanTent:2003mn,Kaiser:2013sna,
Reuter:2015rta,Karamitsos:2017elm}}. The form of~\eqref{eq:geodesic} shows that this trajectory is a geodesic. Thus, the theory described by~\eqref{eq:free lagrangian} is equivalent to that of a free particle moving in field space. Note that we are free to scale and shift our affine parameter so that ${t\neq u}$. This is simply a manifestation of the fact that the Lagrangian~\eqref{eq:free lagrangian} is invariant under shifts of time and its EoMs~\eqref{eq:geodesic} are invariant under rescalings of time.

We now add a potential term to~\eqref{eq:free lagrangian} so that the Lagrangian becomes
\begin{equation}\label{eq:force lagrangian}
{\cal L}\ =\ \hf k_{ij}(\bm{\varphi})\;\dot{\varphi}^i\dot{\varphi}^j\: -\: V(\bm{\varphi})\;.
\end{equation}
The EoMs derived from this Lagrangian are
\begin{equation}\label{eq:force EOM}
\ddot{\varphi}^i\: +\: \Gamma^i_{jk}\,\dot{\varphi}^j\dot{\varphi}^k\ =\ -\, k^{ij}V_{,j}\;.
\end{equation}
Observe that there is now a term on the RHS, which can be interpreted as an external conservative force acting on the particle in field space.

The question that naturally arises is whether we can use the Eisenhart lift to construct a higher-dimensional space equipped with a metric such that the geodesic equations reduce to the EoMs given in \eqref{eq:force EOM}. This is indeed possible by following a completely analogous procedure to the one for the classical particle in Section~\ref{sec:classical}. We add a new coordinate to our field space, which corresponds to adding a new \emph{fictitious} scalar field~$\chi$ to our theory. This leads to the following Lagrangian:
\begin{equation}\label{eq:new lagrangian}
{\cal L}\ =\ \hf k_{ij}(\bm{\varphi})\;\dot{\varphi}^i\dot{\varphi}^j\: +\: \hf \frac{M^4}{V(\bm{\varphi})}\,\dot{\chi}^2\;,
\end{equation}
where $M$ is again an arbitrary mass scale. We incorporate the field $\chi$ into the new $(N\!+\!1)$--dimensional field space by introducing $\phi^A$ with an index $A$ that runs from~$1$ to $N\!+\!1$, with $\phi^i \equiv \varphi^i$ for $1\leq i\leq N$, and $\phi^{N+1}\equiv\chi$. We also define an extended version of the field space metric:
\begin{equation}
G_{AB}\ \equiv\ 
\begin{pmatrix}  
k_{ij}	&	0\\
0		&	\dfrac{M^4}{V}
\end{pmatrix}.
\end{equation}
Note that $G_{AB}$ does not depend on the fictitious field~$\chi$. With these definitions, the Lagrangian~\eqref{eq:new lagrangian} can be rewritten as
\begin{equation}\label{eq:general 1d lagrangian}
{\cal L}\ =\ \hf G_{AB}\,  \dot{\phi}^A  \dot{\phi}^B\; .
\end{equation}
This Lagrangian is of the same form as~\eqref{eq:free lagrangian}. Thus, we expect the evolution of the fields governed by~\eqref{eq:general 1d lagrangian} to be equivalent to a free particle moving in an $(N\!+\!1)$--dimensional field space equipped with the metric $G_{AB}$.

We now explicitly calculate the evolution of the extended system. The EoM  for the new field $\chi$ (or equivalently~$\phi^{N+1}$) is
\begin{equation}
\frac{d}{dt}\fb{\dot{\chi}}{V(\bm{\varphi})}\ =\ 0\;.
\end{equation}
Consequently, $\chi$ must satisfy
\begin{equation}
\dot{\chi}\ =\ A \frac{V(\bm{\varphi})}{M^2}\ ,\label{eq:chidot sol}
\end{equation}
where $A$ is a constant. The EoMs  for $\varphi^i$ read
\begin{equation}\label{eq:varphi eom}
\ddot{\varphi}^i\: +\: \Gamma^i_{jk}\,\dot{\varphi}^j\dot{\varphi}^k\ =\ -\, \hf k^{ij}V_{,j}\,\frac{M^4\dot{\chi}^2}{V^2}\ .
\end{equation}
If $\chi$ satisfies its EoM, then \eqref{eq:varphi eom} becomes
\begin{equation}
\ddot{\varphi}^i\: +\: \Gamma^i_{jk}\dot{\varphi}^j\dot{\varphi}^k\ =\ 
-\,\frac{A^2}{2} k^{ij}V_{,j}\; .\label{eq:varphi A eom}
\end{equation}
If $A^2=2$ (which can be satisfied by selecting specific initial conditions), then \eqref{eq:varphi A eom} is identical to~\eqref{eq:force EOM}. Therefore, the fields~$\varphi^i$ will evolve in the same way as they did when governed by the Lagrangian~\eqref{eq:force lagrangian}. As before, the free parameter $A$ arises from our freedom in parametrising the geodesics. Choosing $A^2\neq2$ causes the system to evolve either in slow motion ($A^2<2$) or fast-forward ($A^2>2$).

\section{Noether Symmetries and Killing's Equation}
\label{sec:Killing}
We now wish to investigate whether~\eqref{eq:general 1d lagrangian} admits any symmetries and what equations these symmetries must satisfy. For similar considerations in different settings, see~\cite{deRitis:1990ba,Capozziello:1993vr,Capozziello:1996bi,Kaewkhao:2017evn}. To start with, we consider a transformation of the fields:
\begin{equation}
\phi^A\ \rightarrow\ \phi^{\prime A}\ =\ \phi^A\: +\: \xi^A\;.\label{eq:phi transformation}
\end{equation}
Under this transformation, the change in the Lagrangian is
\begin{equation}
\delta {\cal L}\ =\ \left(G_{AB}\, \xi^A_{,C}\: +\: \hf G_{BC,A} \,\xi^A\right)\dot{\phi}^B\dot{\phi}^C\;.
\end{equation}
For this to be a true symmetry of the theory, $\delta \mathcal{L}$ must vanish regardless of the field configuration. Thus, each coefficient of $\dot{\phi}^B\dot{\phi}^C$ must vanish separately and the transformations $\xi^A$ must satisfy the relation
\begin{equation}
G_{AB}\xi^A_{,C}\: +\: G_{AC}\xi^A_{,B}\: +\:  G_{BC,A}\xi^A\ =\ 0\;.
\end{equation}
This can be recast in the form
\begin{equation}
\label{eq:Killing}
\nabla_B\xi_C\: +\: \nabla_C\xi_B\ =\ 0\;,
\end{equation}
where $\nabla_A \xi_B=\partial_A\xi_B-\Gamma^C_{AB}\xi_C$ is the covariant derivative\- in the extended field space, and we have defined ${\xi_A \equiv G_{AB}\xi^B}$. Equation~\eqref{eq:Killing} is Killing's equation for the field space metric $G_{AB}$. We therefore see that the Noether symmetries of the theory are isometries of the field space described by Killing vectors and vice versa.

Our aim is now to find the symmetries that do not involve the fictitious field $\chi$. To this end, we set ${\xi^{N+1}=\xi^A_{,N+1}=0}$. In this case, the Killing equations~\eqref{eq:Killing} reduce to
\begin{equation}
\nabla_i\xi_j\: +\: \nabla_j\xi_i\ =\ 0\;,\hspace{2em}
\xi^i V_{,i}\ =\ 0\;,\label{eq:Killing original}
\end{equation}
where $1\leq i,j\leq N$. Observe that the second of these equations comes from setting ${B=C=N\!+\!1}$ in \eqref{eq:Killing}. These are precisely the conditions that must be satisfied by the Noether symmetries of the Lagrangian~\eqref{eq:force lagrangian}.

\section{Generalisation to Four Dimensional Field Theories}
\label{sec:4d}
In four (and higher) dimensions, the Lagrangian for $N$ scalar fields~$\varphi^i(t,{\bf x})$ is
\begin{equation}
{\cal L}\ =\ \sqrt{-g}\left(\hf g^{\mu\nu}k_{ij}(\bm{\varphi})\;\partial_\mu\varphi^i\partial_\nu\varphi^j-V(\bm{\varphi})\right),\label{eq:4d lagrangian}
\end{equation}
where $g_{\mu\nu}$ is the spacetime metric, with $g\equiv \det g_{\mu\nu}$. As we will show, the dependence of the fields $\varphi^i$ on the spatial dimensions~${\bf x}$ makes it more difficult to reproduce the effects of the potential with a curved field space. For example, we may be tempted to extend the Lagrangian in the following way:
\begin{equation}
{\cal L}=\sqrt{-g}\left(\hf g^{\mu\nu}k_{ij}(\bm{\varphi})\;\partial_\mu\varphi^i\partial_\nu\varphi^j+\hf\frac{M^4}{V(\bm{\varphi})}g^{\mu\nu}\partial_\mu\chi\partial_\nu\chi\right).\label{eq:4d Lag}
\end{equation}
Varying ${\cal L}$ with respect to $\chi$ gives the following EoM:
\begin{equation}
\nabla_\mu A^\mu\ =\ 0\;,\label{eq:4d A_mu}
\end{equation}
where $A^\mu=M^2 \partial_\mu\chi/V$. If $A^\mu$ satisfies $A_\mu A^\mu=2$, then we recover the original EoMs as derived from the Lagrangian~\eqref{eq:4d lagrangian}. However, because of the sum over~$\mu$,~\eqref{eq:4d A_mu}~does \emph{not} imply that $A^\mu$ is a constant, even for flat spacetime, and so this condition is not met in general. Hence, the Lagrangian~\eqref{eq:4d Lag} is not a valid generalisation of the Eisenhart-lift formalism.

However, we find that it is still possible to construct a purely kinetic Lagrangian that reproduces the EoMs  of~\eqref{eq:4d lagrangian} by introducing a vector field $B^\mu$ instead. Consider the following Lagrangian:
\begin{equation}
{\cal L}=\sqrt{-g}\left(\hf g^{\mu\nu}k_{ij}(\bm{\varphi})\;\partial_\mu\varphi^i\partial_\nu\varphi^j+\hf \frac{M^4}{V(\bm{\varphi})} \nabla_\mu B^\mu\nabla_\nu B^\nu\right).
\end{equation}
The EoM for $B^\mu$ is
\begin{equation}
\partial_\mu\fb{\nabla_\nu B^\nu}{V(\bm{\varphi})}\ =\ 0\;,
\end{equation}
which \emph{does} imply $\nabla_\nu B^\nu/V$ is a constant. Hence, the EoMs  for the scalar fields $\varphi^i$ are the same as the ones resulting from the Lagrangian~\eqref{eq:4d lagrangian}, up to a global scaling of the coordinates in a way similar to the rescaling of time in one dimension.

In this formulation, the field space and spacetime are inextricably linked. The fictitious fields $B^\mu$ are an integral part of the extended field space, yet carry a spacetime index $\mu$ and thus transform as a Lorentz four-vector. This link can be made manifest by defining a mixed vierbein~\cite{Schwinger:1963re}~$e_m^\mu$ such that
\begin{equation}
B^\mu\ \equiv\ e_m^\mu B^m\;.
\end{equation}
Here $m$ is a field space index, with ${N+1\leq m\leq N+4}$. We take $e^\mu_m$ to satisfy the vierbein conditions:
\begin{equation}
e^\mu_m e^\nu_n\eta^{mn}=g^{\mu\nu}\:,\hspace{1em}
e^\mu_m e^\nu_n g_{\mu\nu} = \eta_{mn}\:,\hspace{1em}
\nabla_\nu e^\mu_m =0\:.
\end{equation}
This allows us to generalise the Lagrangian~\eqref{eq:general 1d lagrangian} to four dimensions${}$ as follows:
\begin{equation}
{\cal L}\ =\ \hf \sqrt{-g}\, H_{AB}^{\mu\nu} \, \partial_\mu\phi^A \partial_\nu\phi^B,\label{eq:4d h lagrangian}
\end{equation}
where $A$ and $B$ run from 1 to $N+4$. In this notation, $\phi^i\equiv \varphi^i$ for $1\leq i\leq N$, $\phi^m\equiv B^{m}$ for $N+1\leq m\leq N+4$, and $H_{AB}^{\mu\nu}$ is given by
\begin{equation}
H_{AB}^{\mu\nu}\ =\ \begin{pmatrix}
g^{\mu\nu}k_{ij}&0\\
0&\dfrac{M^4}{V}e^\mu_{m}e^\nu_{n}
\end{pmatrix}.
\end{equation}
There is a fundamental difference between this result and the one obtained in the previous formulation for one-dimensional fields. In one dimension, we were able to recreate the effects of a potential by simply extending the field space. However, in four dimensions, we must fundamentally alter the form of the kinetic terms, since~$H_{AB}^{\mu\nu}$ cannot be factorised into a spacetime metric $g^{\mu\nu}$ and a field space metric $G_{AB}$, i.e. $H^{\mu\nu}_{AB}\neq g^{\mu\nu}G_{AB}$.

Let us explore the conditions under which Lagrangian${}$~\eqref{eq:4d h lagrangian} admits Noether symmetries and  see whether we can still retain the connection to isometries of the field space. Under transformation~\eqref{eq:phi transformation}, the Lagrangian~\eqref{eq:4d h lagrangian} changes by 
\begin{equation}
\label{eq:Lvar}
\delta {\cal L}\ =\ \sqrt{-g} \left(H^{\mu\nu}_{AB} \, \xi^A_{,C}+\hf H^{\mu\nu}_{BC,A} \, \xi^A\right)\partial_\mu\phi^B\partial_\nu\phi^C.
\end{equation}
In order for this transformation to be a true symmetry, the variation \eqref{eq:Lvar} must vanish regardless of the field configuration and so each term must be set to zero individually. This requirement yields the following equations:
\begin{equation}
H^{\mu\nu}_{AB}\xi^A_{,C}\: +\: \Gamma^{\mu\nu}_{CBA}\xi^A\:
+\: H^{\mu\nu}_{AC}\xi^A_{,B}\: +\: \Gamma^{\mu\nu}_{BCA}\xi^A\ =\ 0\;,\label{eq:H Killing}
\end{equation}
where
\begin{equation}
\Gamma^{\mu\nu}_{ABC}\ \equiv\ \hf \left(H^{\mu\nu}_{AB,C}\: +\: H^{\mu\nu}_{AC,B}\: -\: H^{\mu\nu}_{BC,A}\right).
\end{equation}
If we treat~$H^{\mu\nu}_{AB}$ as ten different field-space metrics (one for each symmetric combination of $\mu$ and $\nu$), then~\eqref{eq:H Killing} becomes a set of ten Killing's equations, one for each metric. We again look for symmetries that do not involve the fictitious fields~$B^m$. We therefore set ${\xi^{m}=\xi^A_{,m}=0}$ with ${N+1\leq m\leq N+4}$. We find that the equations~\eqref{eq:H Killing} reduce to~\eqref{eq:Killing original} as before. These are the conditions that must be satisfied by the Noether symmetries of the Lagrangian~\eqref{eq:4d lagrangian}.

Finally, we consider the EoMs deduced from~\eqref{eq:4d h lagrangian} and compare them with the geodesic equation~\eqref{eq:geodesic u}. Varying the Lagrangian~\eqref{eq:4d h lagrangian} with respect to the field $\phi^A$ yields
\begin{equation}
\label{eq:gen geodesic}
H^{\mu\nu}_{AB}\nabla_\mu\nabla_\nu\phi^B\: +\: \Gamma^{\mu\nu}_{ABC}\,\partial_\mu\phi^B\partial_\nu\phi^C\ =\ 0\;.
\end{equation}
When the fields depend on only one coordinate, \eqref{eq:gen geodesic} reduces to the geodesic equation~\eqref{eq:geodesic u}. Solving \eqref{eq:gen geodesic} in this case allows us to determine the \emph{world-line} of a single point evolving in the extended field space, enabling us to recover the results of the homogeneous case.

In general, the fields will depend on $D$ coordinates, so Equation~\eqref{eq:gen geodesic} describes a \mbox{$D$-dimensional} \emph{world-volume} that arises as a result of an object with $(D\!-\!1)$ spatial dimensions evolving in an \mbox{$(N\!+\!D)$-dimensional} bulk space. Furthermore, the Lagrangian \eqref{eq:4d h lagrangian} resembles that of a non-linear sigma model, an example of which is the Polyakov string action~\cite{Brink:1976sc,Polyakov:1981rd}. Non-linear sigma models\- describe objects that extremise their world-volume while moving in a bulk. However, these models require that~$H^{\mu\nu}_{AB}$ be separable into a world-sheet metric${}$~$g^{\mu\nu}$ and a bulk metric $G_{AB}$. This is something that cannot be realised in our extension of the Eisenhart-lift formalism. Consequently, the formalism presented in this section differs from the one followed in non-linear sigma models and should therefore be regarded as their generalisation${}$.

\section{Discussion}
\label{sec:discuss}
We have extended the Eisenhart-lift formalism to field space by introducing an additional degree of freedom which can replicate the effects of a potential in a differential-geometric manner. We have thus shown that there is a one-to-one correspondence between a field theory with a potential and a free theory with one extra degree of freedom and intrinsic curvature. This degree of freedom is dynamical, unlike an auxiliary field, and hence can be better described as \emph{fictitious}. We have found that the Noether symmetries of the original theory with a potential are equivalent to the isometries of the extended field space described by Killing vectors. In this field space, the system evolves along a geodesic.

Our extension of the Eisenhart lift can be applied to a variety of situations. One example is multifield inflation~\cite{Sasaki:1998ug,Gordon:2000hv,Seery:2005gb,Arroja:2008yy,Langlois:2008qf,
GrootNibbelink:2000vx,GrootNibbelink:2001qt, vanTent:2003mn,Kaiser:2013sna, Reuter:2015rta,Karamitsos:2017elm,Gong:2002cx,Langlois:2008mn, Gong:2011uw,Copeland:1994vg,GomezReino:2006wv,Elliston:2012ab}, where the multiple scalar fields are homogeneous at leading order. Thus, the relevant Lagrangian is~\eqref{eq:general 1d lagrangian} and we can use our method to describe the trajectory as a geodesic in the new extended field space. One may now ask how the slow-roll regime $ \dot\varphi^2 \ll V$ may be identified, since we no longer have a potential. However,~\eqref{eq:chidot sol} implies that at the classical level, this condition is equivalent to $\dot \varphi^{2}\ll M^2\dot \chi$. This condition is independent of the choice of the mass scale~$M$, since the value of~$\dot{\chi}$ required to recover the original EoMs~\eqref{eq:force EOM} scales as $M^{-2}$, as can be easily seen from~\eqref{eq:chidot sol}. In our formalism, the slow-roll condition is therefore replaced with a slow-roll \emph{hierarchy} between the fields that is satisfied when the system is evolving faster in the fictitious direction than in any other.

Our approach may also give some insight into the initial conditions problem of inflation \cite{Linde:2017pwt,Ijjas:2013vea}. Since we have now encoded the entire theory into the structure of the field space, initial conditions can be studied without reference to the inflationary potential. Instead, we may study the problem in terms of a measure dependent purely on the geometric structure of the extended field-space manifold.

In order to deal with non-homogeneous perturbations, we must extend the Eisenhart lift to field theories in four dimensions. We have outlined how this may be achieved in Section~\ref{sec:4d}. In this case, a non-trivial link between spacetime and the field space must be considered. We can still describe the system geometrically as a 3-brane moving in an $(N\!+\!4)$--dimensional bulk. However, this link implies that the system does not evolve as a classical 3-brane in the extended field space. Instead, its motion is governed by Equation~\eqref{eq:gen geodesic}. This should be contrasted with the homogeneous case, where the system\- did evolve as a classical particle in the extended field space.

In this paper, we have restricted our attention to classical\- field theories. An obvious next step would be to examine how a theory with a potential is related to the purely-kinetic extended theory after quantisation. In~this~context, it is worth noting that our extended theory\- does not contain the dimension-2 term $m^2\phi^2$ that are responsible\- for the gauge hierarchy problem \cite{Gildener:1976ai,Weinberg:1978ym}, as these are absorbed into the geometry of the extended manifold. Our extended approach therefore offers a novel avenue to investigate this problem.

Evidently, the Eisenhart lift treats conservative forces in our conventional world as projections like {\em shadows} that emanate from another higher dimensional cosmos. For instance, this feature is beautifully illustrated for the harmonic oscillator in~Figure~\ref{fig:SHM}, where we can only perceive the projection~$x(t)$, but not the other dimension~$y(t)$. Hence, the formalism presented in this paper\- seems to provide a simple realisation of Plato's world of shadows reminiscent\- to a cave which one hopes to escape from and so gain a deeper understanding of the fundamental laws of nature.

\begin{acknowledgements}
The authors would like to thank Jack Holguin and Mari\'{a}n Fecko for useful comments. KF is supported by the University of Manchester through the President's Doctoral Scholar Award. The work of SK was partially supported by an STFC PhD studentship.  The work of AP is supported by the Lancaster--Manchester--Sheffield Consortium for Fundamental Physics under STFC research grant ST/L000520/1.
\end{acknowledgements}

\end{document}